\documentclass[11pt]{article}

\usepackage{a4}
\usepackage[latin1]{inputenc}
\usepackage[all]{xy}
\usepackage{fullpage}
\usepackage{marvosym}
\usepackage{graphicx}
\usepackage{fancybox}
\usepackage{natbib}

\newcommand{\Al}{\mbox{\footnotesize \textsf{A}}}
\newcommand{\Be}{\mbox{\footnotesize \textsf{B}}}

\newcommand{\euro}{\textrm{~\EUR}}

\title{Bubbles are rational}
\author{Pierre Lescanne\\
Université de Lyon, École normale supérieure de Lyon, CNRS (LIP), \\ 46 all\'ee
d'Italie, 69364 Lyon, France}

\begin{document}
\maketitle

  \hrule

\begin{abstract}

  As we show using the notion of equilibrium in the theory of infinite sequential games,
  bubbles and escalations are rational for economic and environmental agents, who believe
  in an infinite world. This goes against a vision of a self regulating, wise and pacific
  economy in equilibrium. In other words, in this context, equilibrium is not a synonymous
  of stability.  We attempt to draw from this statement methodological consequences and a
  new approach to economics. To the mindware of economic agents (a concept due to
  cognitive psychology) we propose to add coinduction to properly reason on infinite
  games. This way we refine the notion of rationality.

\medskip

\end{abstract}

\hrule

\bigskip

\hfill\parbox{9cm}{
There is always some madness in love. But there is also always some reason in madness.

\rightline{\emph{Friedrich Nietzsche}}
\rightline{\textsf{Thus Spoke Zarathustra} (1898)}
}

\bigskip

Traders speculate with no limit, countries bankrupt, world consumes energy like crazy.
Are our models of a self-regulating economy, which rejects bubbles as unlikely, adapted?  Are we
sure to understand how agents act?

The 2008 subprime crisis has shed light on two problems.  One, the crisis is not due to mad
actors, but to the interaction of intelligent actors.  Two, the current tools
of economics, based on concepts elaborated in the middle of the previous century are out
of date. 

Here we focus on a well known phenomenon, namely escalation\footnote{A well known
  phenomenon in the literature from Macbeth to Madame Bovary} whose rationality has been
questioned and even refuted as paradoxical.  Escalation consists in taking with no limit a
sequence of decisions with heavier and heavier consequences.  This headlong run strikes
today the economy, the finance and the social and environmental development and is a
characteristic of financial bubbles.  This apparent irrationality has been illustrated by
Newton after the emergence of one of the first financial crisis, namely the South Sea
Bubble when he said that ``he could calculate the motions of erratic bodies, but not the
madness of a multitude''.  In escalation, agents behave absolutely rationally, provided
they believe in the endless availability of natural or financial resources.  Indeed the
trader or the investor is rational because he reasons in his own world which he thinks
infinite, because he aims at maximizing his profits and because he believes that he can create money with no limit.
Amazingly a person involved in an escalation can bid indefinitely with no respect for his
loss.  For an external observer, the decision taker implied in an escalation seems to have
lost his common sense, but from the point of view of the decision taker himself locked in his
closed world he is perfectly rational.  The consistence of this attitude will be proved by
a subtle and correct reasoning on infiniteness.  This level dependent perception is
probably what distinguishes instrumental rationality from epistemic rationality (see
Section~\ref{sec:esc-psy}).  Hence, an agent who captures both internal and external
vision and takes both into account is\emph{ epistemicly rational} whereas an agent
prisoner of the system is only \emph{instrumentally rational}.  The first agent will
foresee the bubble outburst, when the second agent will remain blind.

At last, we do not say that an escalation followed by a collapse which can be its consequence
is unavoidable, but we claim that it is plausible since escalation is supported by a
rational behavior.

\section{Toward new tools for analyzing systems} 
\hfill\parbox{12cm}{\begin{quote}

It is human nature to think wisely and to act absurdly

\rightline{\emph{Anatole France}} \rightline{\textsf{Le livre de mon ami}}
\rightline{\textit{\textsf{The book of my friend}}}
\end{quote}}

\bigskip

In this article, we study systems with agents, where a system is an organization in which
elementary entities able to reason are called ``agents''.  In what follows, we use also
the term ``player'', by analogy with games which we will use as our paradigm.  According
to its concern, an agent takes decisions which are consequences of her preference or of
her appreciation of what she can gain.  Of course, those choices influence the global
behavior of the system. A behavior, we are interested in, is an \emph{equilibrium}, in
which the decision of the agents are made in order to maximise their returns.  In that
sense, equilibrium is synonymous of \emph{stability}. But as we will show, this can
coincide with a quick evolution of the main parameters of the system,\footnote{In general
  relativity, a black hole is the result of an equilibrium, whereas we know how extreme
  its behavior is.} as for instance the quick raising or decreasing of a price.  Thus an
equilibrium can yield a big instability of the main parameters of the system, as this is
the case in escalation, a concept we will focus on.  Roughly speaking we investigate
sequences \emph{equilibrium-decision, equilibrium-decision}, etc. and instability can
result as an outcome of these sequences.  In infinite games, equilibrium is no more
synonymous of stability.  This propensity of the agents toward optimization is what people
call ``rationality'', in other words agents are gifted with a reason which they use for
their advantage in their decisions.  But rationality has two faces, according to how it is
viewed. Indeed it can be viewed from inside the system or from outside, that is from a
local or from a global perspective.  These two points of view lead to two opposite
statements.  Escalation which is specifically irrational from a holistic point of view is
rational from a reductionist one.  This phenomenon raises many paradoxes.  First a local
rationality, this of the agents taken individually, can result in a total irrationality
when the system is taken as a whole.  Since the agent is squeezed in her world, it will be
difficult for an external observer to convince her of her mistake.  When a observer
affirms the rationality or the irrationality of an agent he should tell at which level he
stays. Second, a system founded on equilibria can be chaotic and a outburst can be a
consequence of a sequence of equilibria.  Therefore, the expression ``agents in
equilibrium when taking decision'' does not mean ``a system with a slow and regular
evolution''.

Chaos and escalation are not well-understood in scientific disciplines (among others
economics) grounding part of their mathematical model on multi-agents systems.  Therefore, we feel very
close to Jean-Philippe Bouchaud who claims that ``Economics needs a scientific revolution''
\footnote{J-Ph. Bouchaud.  \textsf{Economics needs a scientific revolution}.
  \emph{Nature}, 455:\penalty0 1181, oct 2008.} or David Colander who announces the end of
the economics as we know it and calls for a refoundation.  New conceptual tools should be
used, based on logic as developed by another science of large systems, namely computer science,
on top of which coinduction and coalgebras lie.

\section{A case study: McDonald's Restaurants vs Morris \& Steel}

\hfill\parbox{9cm}{
 Tell me what you eat, I tell you who you are.

\rightline{\emph{Anthelme Brillat-Savarin,}} 

\rightline{\textsf{La physiologie du goût (The physiology of Taste)}}
}

\bigskip

An interesting case of escalation is this which opposed McDonald's Corporation to two
English environmental activists, David Morris and Helen Steel\,\footnote{John Vidal,
  \textsf{McLibel, Burger Culture on Trial}, \emph{Macmillan Publishers}, 1997} in the
infamous so-called \emph{McLibel Trial}, also known as \emph{McDonald's Restaurants v
  Morris \& Steel}. Clearly McDonald's management has a well-thought-out policy, knows how
to argue and takes advice from the best lawyers.  The trial which latest ten years was the
longest-running case in the English history.  The facts started by a small leaflet
campaign, making allegations, not all asserted.  Helped by the stars of the bar
association and firmly grounded on a supposed ``credible threat'' (see
Section~\ref{sec:LescEstRati}), namely that the powerful MacDonald's scares the
activists and can support an endless case, the company suited the activists for libel.
Two of them, David Morris et Helen Steel, decided to defend their case.  At the end of a
very long trial, the judge ruled mostly against the defendants, because some allegations
of the pamphlet could not be proved.  But in retrospect the company lost.  Specifically,
the judge ruled that McDonald's endangered the health of their workers and customers by
"misleading advertising", that they "exploit children", that they were "culpably
responsible" in the infliction of unnecessary cruelty to animals, and that they were
"antipathetic" to unionization and paid their workers low wages.  This was evidenced in
court by a media circus. Eventually, the defendants took the case to the European Court of
Human Rights who ruled against the British government because UK laws had failed to
protect the public right to criticize corporations whose business practices affect
people's lives and the environment; they also ruled that the trial was biased because of
the defendants' comparative lack of resources and what they believed were complex and
oppressive UK libel laws.  Eventually the law was changed.  It has been estimated that
McDonald's Company has lost 10~000~000~£ with further consequence, when the activists
spend 30~000~£, supported by subscription and the compensation ruled by the European Court
of Human Rights.  This is a typical case of escalation.  Although the company saw that it
was trapped and lost in image, it could not be defeated in the court and kept arguing.  It
is interesting to notice that a preliminary debate at the court was on the rationality of
the decision process. The question was whether a judge or a jury would take the case.  The
proposal of a jury was rejected because it was considered more emotional whereas a judge
was considered more rational.

\section{Equilibria in games}
\label{sec:equilibria-games}

In all escalation processes, there is an interaction--competition mechanism and
competition prevails.  Early philosophers noticed that the rules that govern group activities are
those of games, leading to the development of \emph{game theory}.  Indeed like in an actual game,
players cooperate more or less, but overall they act for their own interest.  Early on, the concept of
game  describes how the interaction between the actors of  a system works.   This
applies particularly to economics.   The founding act of game theory is John von
Neumann and Oskar Morgenstern book, \emph{Theory of Games and Economic Behavior},
published in 1944.\footnote{The first paper on game theory is usually attributed to Ernst
  Zermelo and entitled \emph{``Über eine Anwendung der Mengenlehre auf die Theorie des
  Schachspiels'}', in the Proceedings of the Fifth International Congress of Mathematicians
  (1913).}

Jean-Jacques Rousseau tells in his \emph{Discourse on the Origin and Basis of Inequality
  Among Men} (1775) how even acculturated men  act collectively mixing interaction and
selfishness. This leads them to choose an option fitting with their best immediate
interest. He chose the example of a deer hunt which remains infamous. 

\begin{quote}
  ``In this manner, men may have insensibly acquired some gross ideas of mutual
  undertakings, and of the advantages of fulfilling them: that is, just so far as their
  present and apparent interest was concerned: for they were perfect strangers to
  foresight, and were so far from troubling themselves about the distant future, that they
  hardly thought of the morrow. If a deer was to be taken, every one saw that, in order to
  succeed, he must abide faithfully by his post: but if a hare happened to come within the
  reach of any one of them, it is not to be doubted that he pursued it without scruple,
  and, having seized his prey, cared very little, if by so doing he caused his companions
  to miss theirs.''
\end{quote}

In this duality collective -- individual involvement, the  agent adopts a course of action from
which she has no interest to deviate.  A strategic position, where individual behaviors
are stuck because agents do not change their choices is called an
\emph{equilibrium}.    In his book \emph{Researches on the Mathematical Principles of the
  Theory of Wealth} (1838),  the economist and mathematician Antoine-Augustin Cournot
evidenced in the case of duopoly this notion of equilibrium which has not yet its name.
Two companies compete for production of the same objects and try to adjust their production to
optimize their profits.  A Cournot equilibrium is the optimum
quantity which companies must manufacture to earn the most money. One of the
characteristics of Cournot duopolies and Rousseau hunters is the interaction without
cooperation.  We will therefore focus on non cooperative games, in which each player is
selfish and does not attempt to help the others even thought this may help her in the
long term.  Keeping the framework of non cooperative games, we go from Antoine~A. Cournot
to John F.~Nash, who in 1947 stated a general form of equilibrium which one calls \emph{Nash
equilibrium} and which started an active research on non cooperative games.  These games
are defined on a particular form of games which we call \emph{normal form games}.  They
are one shot game where payoff are immediately distributed.   A typical example of
normal form game is the game \emph{stone-paper-scissors}.  Two players show at the same
time one of the three following objects, a stone, a paper or scissors. The rules are as
follows:
\begin{itemize}
\item paper beats stone,
\item stone beats scissors,
\item scissors beat papers.
\end{itemize}
In this game each player must randomize her choices and her optimum strategy is to play
$1/3$ paper, $1/3$ stone and $1/3$ scissors.  The game \emph{matching pennies} is a
simplified version of the game  \emph{stone-paper-scissors}.  Two players whom we  call
\textsf{Alice} and \textsf{Bob} display at the same time a coin, \emph{head} or \emph{tail}. 
\begin{itemize}
\item If both players display \emph{heads} or both players display \emph{tails},
  \textsf{Alice} wins.
\item If both players display mismatching pennies, \textsf{Bob} wins.
\end{itemize}

Again players must play head and tail with equal probability.  We will not extend on this
type of games, because there is too much emphasis given to the value of the payoff, which
is at the core of probability and we claim that the actors play without quantifying their
gains.  We are interested by sequential games with no probability, as proposed by Harold
Kuhn in 1953, following Zermelo. In what follows values attributed to players will be
symbolic.  No computations are required, only comparisons.

\section{Sequential games}
\label{sec:sequential-games}

In a sequential game, players play several runs, each player at her turn. The distribution
of the payoffs takes place at the end of the game.  In these games, there are many kinds
of equilibria, but those we are interested in are the so-called \emph{backward
  equilibria}, since they translate the rationality of the choices of the players.  They
are called ``backward'' since they compute backward from the end of the game.  Without
loss of generality, we consider only two player games. For instance, consider a variant of
the game \emph{matching pennies}, where this time, players \textsf{Alice} and
\textsf{Bob} play one after the other.  More precisely, \textsf{Alice} plays first, then
\textsf{Bob}, then \textsf{Alice} again.  In case of a matching, two heads in a row or two
tails in a row, \textsf{Alice} wins the set. In the other case \textsf{Alice} looses her sets.
The scores are accumulated and counted at the end of the game.  We count the global win,
loss or tie which corresponds respectively as \emph{w}, $\ell$ or \emph{t}.  For instance,
if \textsf{Alice} plays head, then \textsf{Bob} plays tail, and \textsf{Alice} plays head,
there is no matching, then \textsf{Alice} globally looses and \textsf{Bob} globally wins,
resulting in $\ell$.  If \textsf{Alice} plays head, \textsf{Bob} plays tail and
\textsf{Alice} plays tail, there is a global tie for both resulting in \emph{t}.

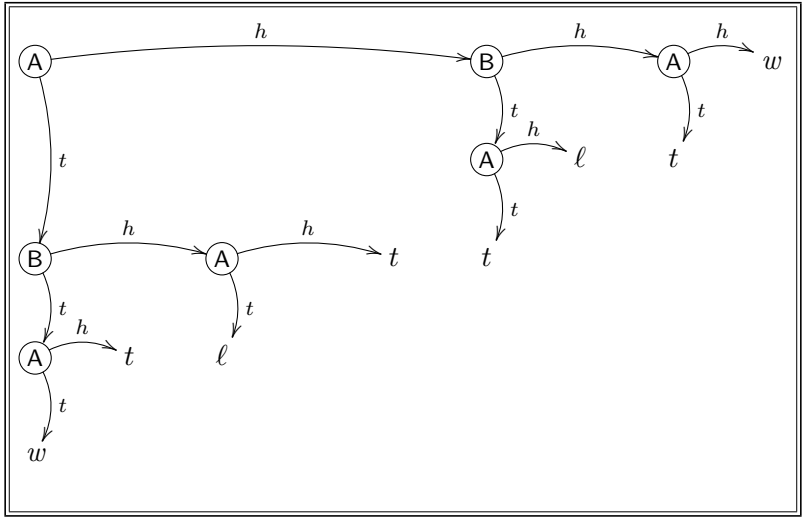
\begin{figure}[ht!]
  \centering \doublebox{\parbox{.6\textwidth}{
\begin{center}
    \xymatrix{ 
      *+[o][F]{\Al} \ar@/^/[rrrrr]^h \ar@/^/[dd]^t &&&&& *+[o][F]{\Be}
      \ar@/^/[rr]^h \ar@/^/[d]^t &&*+[o][F]{\Al} \ar@/^/[d]^t\ar@/^/[r]^h &
      \emph{w}\\
      &&&&&*+[o][F]{\Al}\ar@/^/[d]^t\ar@/^/[r]^h&\ell& t\\
      *+[o][F]{\Be}\ar@/^/[rr]^h \ar@/^/[d]^t && *+[o][F]{\Al} \ar@/^/[rr]^h\ar@/^/[d]^t&& t & t &&\\
      *+[o][F]{\Al}\ar@/^/[d]^t \ar@/^/[r]^h & t & \ell\\
      \emph{w} }
  \end{center}
}}
\caption{The matching pennies sequential game, where $\Al$ starts, then $\Be$~plays,
  then $\Al$ plays again.}
\label{fig:appar}
\end{figure}

This games has an almost obvious winning strategy, but we consider it only as an
illustration, in order to show that we must consider even the most stupid strategies,
according to the counterfactuality.\footnote{A
   reasoning is counterfactual if it relies on hypotheses which are not necessarily
   plausible.  ``If there are Martians, they try to communicate with us''.}  Very naturally this
game is presented by the diagram of Figure~\ref{fig:appar} where \textsf{Alice} is identified by
\Al{} and \textsf{Bob} is identified by \Be{} and where arrows are the steps of the
game. Label $h$ of an arrow shows that the player plays \emph{head}, whereas label $t$
shows that player plays \emph{tail}.  Remind that the value of the payoffs has no meaning
but the fact that they can be compared. Thus $0 < 1$ and $1 <2$.  Number $0$ means that
player never won, number $1$ means that player won as often as her opponent and number
$2$ means that player won twice more that her opponent. 

On the diagram, each player has two choices: either to go down (head) or to go right
(tail).  Starting at the leftmost uppermost node, following the arrow labeled $h$, then
the arrow labeled $t$, then the arrow labeled $h$, one gets to the end of the game.  The
tag $\ell$ means ``\textsf{Alice} globally looses and \textsf{Bob} globally wins''.  Let us call
$h\,t\,h$ such a strategy profile.

\begin{figure}[!tp]
  \centering
     \doublebox{\parbox{.6\textwidth}{
   \begin{center}
     \xymatrix{
      *+[o][F]{\Al} \ar@{=>}@/^/[rrrrr]^h \ar@/^/[dd]^t &&&&&
      *+[o][F]{\Be} \ar@/^/[rr]^h \ar@{=>}@/^/[d]^t &&
      *+[o][F]{\Al}\ar@/^/[d]^t\ar@{=>}@/^/[r]^h &
      \emph{w}\\
      &&&&&*+[o][F]{\Al}\ar@{=>}@/^/[d]^t\ar@/^/[r]^h&\ell& t\\
      *+[o][F]{\Be}\ar@{=>}@/^/[rr]^h \ar@/^/[d]^t 
      && *+[o][F]{\Al} \ar@{=>}@/^/[rr]^h\ar@/^/[d]^t&& t & t &&\\
      *+[o][F]{\Al}\ar@{=>}@/^/[d]^t \ar@/^/[r]^h & t & \ell\\
      \emph{w}}
  \end{center}
  \begin{center}
    \xymatrix{
      *+[o][F]{\Al} \ar@/^/[rrrrr]^h \ar@{=>}@/^/[dd]^t &&&&&
      *+[o][F]{\Be} \ar@/^/[rr]^h \ar@{=>}@/^/[d]^t &&
      *+[o][F]{\Al}\ar@/^/[d]^t\ar@{=>}@/^/[r]^h &
      \emph{w}\\
      &&&&&*+[o][F]{\Al}\ar@{=>}@/^/[d]^t\ar@/^/[r]^h&\ell& t\\
      *+[o][F]{\Be}\ar@{=>}@/^/[rr]^h \ar@/^/[d]^t 
      && *+[o][F]{\Al} \ar@{=>}@/^/[rr]^h\ar@/^/[d]^t&& t & t &&\\
      *+[o][F]{\Al}\ar@{=>}@/^/[d]^t \ar@/^/[r]^h & t & \ell\\
      \emph{w}}
  \end{center}
}}
  \caption{The two equilibria of the matching pennies.  \emph{ The  double
      arrows correspond to choices made by players}}
\label{fig:4eq}
\end{figure}
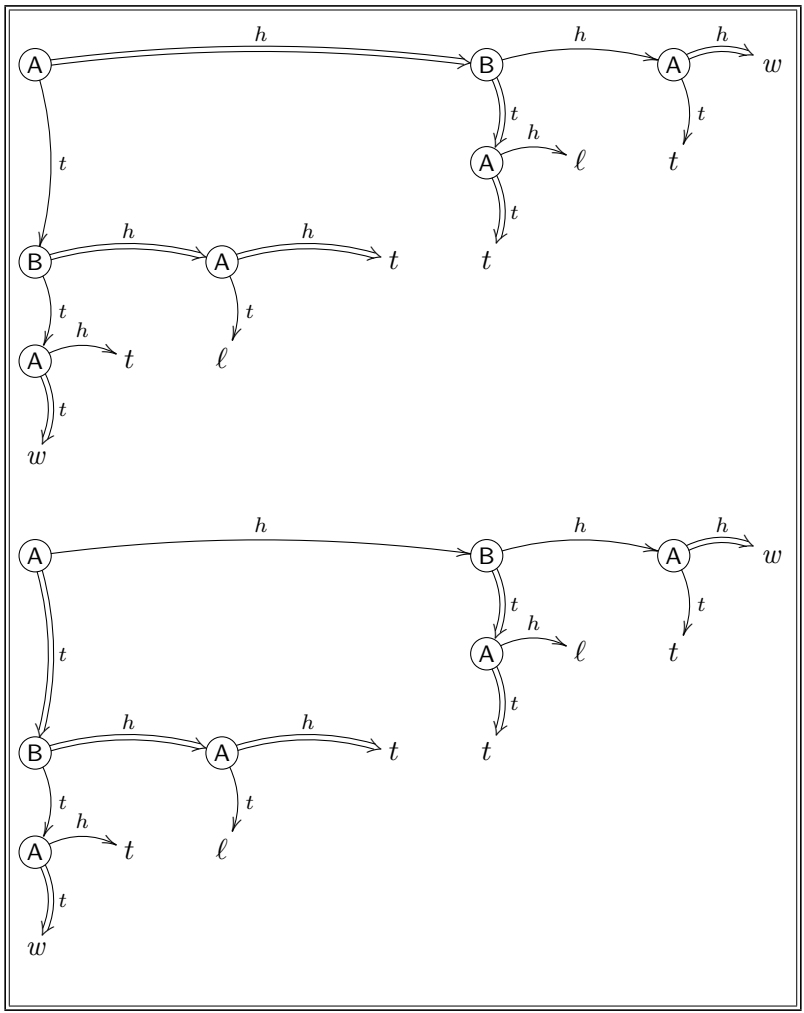

In this game, among the eight strategy profiles, two are equilibria
(Figure~\ref{fig:4eq}).  Thus in the strategy profile $h\,t\,t$ which leads to $t$ if
\textsf{Bob} changes his choice in~$h$, then \textsf{Alice} would chance her choice in $h$
and he would loose, therefore he would not do it. The computation of such an equilibrium
requires what specialists call a \emph{backward induction}.  Let us consider a position in
the game and let us look at positions which come after (following the arrows).  If we
consider this configuration, we see that it is itself a game (a \emph{subgame})
completely included in the game of Figure~\ref{fig:appar}.  The idea is to attribute to
each subgame a couple of payoffs which corresponds to what the equilibrium returns.  To
associate these couples one starts from the end of the game and one goes back to the start
of the game.  Step by step, one builds the assignments of the payoffs for the subgames.
We start from the ends of the game, then we build slightly larger games, then still larger,
to get to the largest game, that is the game we are interested in. In other words, we
build equilibria gradually. At each step, we choose for the player of interest, the larger
payoff and we remember the choice.  The equilibrium is the strategy profile which goes
through the taken positions.

 Characteristics of the analysis of sequential games is \emph{counterfactuality}. We reason in
 all situations as they would be possible.  Clearly, some game positions, located after one
 player has chosen another option will not take place, but they must be taken into account
 to explain how the players reason.  Counterfactuality plays an even more important role in
 infinite games so crucial in escalation.

\paragraph{Which link between backward equilibria and rationality?}
\label{sec:wich-link-between}

Robert Aumann, Nobel Prize winner in economics in 2005, has shown in 1995 that backward
equilibria are rational strategy profiles.  They are strategy profiles in which everybody
knows that nobody can change her choices without loosing or drawing and everybody knows
that rationally no other case can appear.  We say that the players choose these equilibria
from a common knowledge of the rationality of the other players.

\section{0, 1 Games}
\label{sec:0-1-games}

\hfill\parbox{6cm}{\begin{quote}
Vulnerant omnes

Ultima necat

\rightline{\emph{Sentence disposed on sundials,}}
\end{quote}}

\bigskip

We will focus on simple sequential games.  In these games, players have choices to leave
($\ell$) or to continue ($c$). The payoffs are $0\!\euro$ ou~$1\!\euro$\,\footnote{To
  spice up the game, we could replace  $1\!\euro$ by
$1~000~000~000\!\euro$  and therefore say that the players either receive nothing or
receives a billion of euros.  We insist on the fact that the quantity does not count,
only counts the comparison.}
Let us focus on the game with seven players. 

\begin{displaymath}
  \xymatrix{
*++[o][F]{\Al} \ar@/^/[r]^c \ar@/^/[d]^{\ell} &*++[o][F]{\Be} \ar@/^/[r]^c \ar@/^/[d]^{\ell} &*++[o][F]{\Al} \ar@/^/[r]^c \ar@/^/[d]^{\ell} &*++[o][F]{\Be} \ar@/^/[r]^c \ar@/^/[d]^{\ell} &*++[o][F]{\Al} \ar@/^/[r]^c \ar@/^/[d]^{\ell} &*++[o][F]{\Be} \ar@/^/[r]^c \ar@/^/[d]^{\ell} &*++[o][F]{\Al} \ar@/^/[r]^c \ar@/^/[d]^{\ell} &1,0\\
0,1&1,0&0,1&1,0&0,1&1,0&0,1
}
\end{displaymath}

\begin{itemize}
\item At the last turn, \textsf{Alice} must of course choose $c$, since she earns $1\!\euro$.
\item At the penultimate turn, \textsf{Bob} may choose, what a choice!  Indeed, if he
  continues he earns nothing and if he leaves he earns nothing as well.
\item At the fifth turn, \textsf{Alice} continues because in each case, she earns
  $1\!\euro$. And if she leaves she earns nothing at all.
\item At the fourth turn, \textsf{Bob} has the same choice as at the penultimate turn,
  namely to earn nothing or to earn nothing.
\item At the third turn, \textsf{Alice} continues.
\item At the second turn, \textsf{Bob} must choose between nothing or nothing.
\item At the beginning of the game, \textsf{Alice} should do nothing by stop
\end{itemize}

 The $0,1$ game has seven turns and more than one rational strategy profile. They all have
 the characteristics that \textsf{Alice} always continues and \textsf{Bob} does
 whatever he wants.  Let us consider now the $0,1$ game with six turns.

\begin{displaymath}
  \xymatrix{
*++[o][F]{\Al} \ar@/^/[r]^c \ar@/^/[d]^{\ell} &*++[o][F]{\Be} \ar@/^/[r]^c \ar@/^/[d]^{\ell} &*++[o][F]{\Al} \ar@/^/[r]^c \ar@/^/[d]^{\ell} &*++[o][F]{\Be} \ar@/^/[r]^c \ar@/^/[d]^{\ell} &*++[o][F]{\Al} \ar@/^/[r]^c \ar@/^/[d]^{\ell} &*++[o][F]{\Be} \ar@/^/[r]^c \ar@/^/[d]^{\ell} &0,1\\
0,1&1,0&0,1&1,0&0,1&1,0
}
\end{displaymath}

This time the backward induction yields something different. 
\begin{itemize}
\item At the last turn, \textsf{Bob} continues.
\item At the penultimate and fifth turn, \textsf{Alice} does whatever she wants.
\item At the fourth turn, \textsf{Bob} continues.
\item At the third turn, \textsf{Alice} does whatever she wants.
\item Etc.
\end{itemize}
Here is the diagram of the game where \textsf{Alice} leaves always.
\begin{displaymath}
  \xymatrix{
*++[o][F]{\Al} \ar@/^/[r]^c \ar@{=>}@/^/[d]^{\ell} &*++[o][F]{\Be} \ar@{=>}@/^/[r]^c \ar@/^/[d]^{\ell}
&*++[o][F]{\Al} \ar@/^/[r]^c \ar@{=>}@/^/[d]^{\ell} 
&*++[o][F]{\Be} \ar@{=>}@/^/[r]^c \ar@/^/[d]^{\ell} &*++[o][F]{\Al} \ar@/^/[r]^c \ar@{=>}@/^/[d]^{\ell} 
&*++[o][F]{\Be} \ar@{=>}@/^/[r]^c \ar@/^/[d]^{\ell} &0,1\\
0,1&1,0&0,1&1,0&0,1&1,0
}
\end{displaymath}

In this game, and in all $0,1$ games with an even number of turns, the rational strategy
profiles, that are the equilibria, are when \textsf{Bob} continues always and \textsf{Alice
} does whatever she wants.  We will look at infinite $0,1$ games and see what the
equilibria in those games are. 

\section{The dollar auction}

\hfill\parbox{12.7cm}{







\begin{quote}
\hspace*{8cm}{both}

Walk toward each other and the duel starts again

Then by degrees in its dark dementia

The battle intoxicates them; then come to their heart back

This I don't know which god who wants that one is a winner.

\rightline{\emph{Victor Hugo,}} \rightline{\textsf{La Légende des siècles. Le mariage de Roland}}
\end{quote}}

\medskip

\hspace*{8cm}\parbox{8cm}{
\begin{quote}

Honi soit qui mal y pense

\rightline{\emph{King Edward III,}} \rightline{\textsf{Motto of the Order of the Garter}}
\end{quote}}

The dollar auction game has been described by Martin Shubik\footnote{M.~Shubik, \textsf{The
  dollar auction game: A paradox in noncooperative behavior and escalation}. \emph{Journal
    of Conflict Resolution}, 15\penalty0 (1):\penalty0 109--111, 1971.} in 1971 and it is
well known in France under the name \emph{American auction}.  There it is made in some
weddings and it consists in selling the garter of the bride, the products of the sale
helps the couple for their honeymoon or for their settlement.  In his version Shubik sales
through an auction one dollar and the goal is to collect much more than the price of the
object.  The dollar auction consists in selling the object (the dollar) as follows.  Each
time a person bids for $n\euro$, she must put the given amount in a hat and this
amount is never returned to her.  As Shubik wrote it is suggested to wait for starting the auction
that ``the spirits are high'', as it is the case at the end of a wedding party and we can
limit ourself to two bidders, because this does not change the phenomenon.  Actually in
most of the cases, even though it starts with more than two bidders and an increment of $1\euro$,  the auction ends with
two bidders who do not want to give up.  Then one notices an escalation phenomenon. The
bidders are going to pay more than the value of the object and more than what they
wanted to invest at the beginning.  They keep fighting in order to acquire the garter or the
George Washington bill. Actually they have invested so much that they do not want
to give up without getting the desired object.

In Shubik analysis and in this of his successor, we see two contradictory statements:
\begin{itemize}
\item \emph{To have escalation, the game must be infinite.}  More precisely, Shubik writes
  that ``the analysis is confined to a (possibly infinite) game without a termination
  point, as no particular phenomenon occurs if an upper bound is introduced''.
\item \emph{Shubik and his followers analyze finite games}, then extrapolate their results
  to infinite games.
\end{itemize}

From that, they conclude that the only equilibrium is this where no player starts the
auction and never bids.  They claim that the rational player would reject at any price any type of
escalation.  We have shown that this strategy profile is not an equilibrium in the dollar
auction.  To extrapolate finite games to  infinite games, as Shubik and those who have adopted
his reasoning do, is wrong.   We will come back on it later.


\section{The infinite sequential games : back to the $0,1$ game}
\label{sec:back-on-0,1}

\hfill\parbox{10.75cm}{





\begin{quote}
  \begin{it}
    He walks on the immense plain

Goes ahead, comes back, throws the grain further up

Reopens his hand, and starts again,

And I meditate, obscur witness.
  \end{it}

\rightline{\emph{Victor Hugo,}} 
\rightline{\textsf{Saison des semailles. Le soir}}
\end{quote}
}

\bigskip

In $0,1$ games we have limited the number of turns.  But nothing prohibits to consider
infinitely many turns. 

\begin{displaymath}
  \xymatrix{
*++[o][F]{\Al} \ar@/^/[r]^c \ar@/^/[d]^{\ell} &*++[o][F]{\Be} \ar@/^/[r]^c \ar@/^/[d]^{\ell}
&*++[o][F]{\Al} \ar@/^/[r]^c \ar@/^/[d]^{\ell} &*++[o][F]{\Be} \ar@/^/[r]^c \ar@/^/[d]^{\ell} 
&*++[o][F]{\Al} \ar@/^/[r]^c \ar@/^/[d]^{\ell} &*++[o][F]{\Be} \ar@{.>}@/^/[r]^c \ar@/^/[d]^{\ell} 
&\ar@{.>}@/^/[r]^c \ar@{.>}@/^/[d]^{\ell} &\ar@{.>}@/^/[r]^c&\\
0,1&1,0&0,1&1,0&0,1&1,0&&
}
\end{displaymath}

What are the equilibria in this case? One can no more define the equilibria by backward
induction, but we can use a similar method\footnote{The specialists following Selten
(1965) speak of \emph{subgame perfect equilibria}, but we propose to call them \emph{backward
coinduction equilibria}. } which assumes that we reason on infinite objects.  Let us call it
\emph{backward coinduction}.  Actually there are at least two equilibria:
\begin{enumerate}
\item \textsf{Alice} leaves always and \textsf{Bob} continues always (Figure~\ref{fig:AaBc}),
   \begin{figure}[hbt!]
  \centering
  \doublebox{\parbox{.85\textwidth}{ \centering \xymatrix{ *++[o][F]{\Al}
        \ar@{->}@/^/[r]^c \ar@{=>}@/^/[d]^{\ell} &*++[o][F]{\Be} \ar@{=>}@/^/[r]^c
        \ar@{->}@/^/[d]^{\ell} &*++[o][F]{\Al} \ar@{->}@/^/[r]^c \ar@{=>}@/^/[d]^{\ell}
        &*++[o][F]{\Be} \ar@{=>}@/^/[r]^c \ar@{->}@/^/[d]^{\ell} &*++[o][F]{\Al}
        \ar@{->}@/^/[r]^c \ar@{=>}@/^/[d]^{\ell} &*++[o][F]{\Be} \ar@2{.>}@/^/[r]^c
        \ar@{->}@/^/[d]^{\ell} &\ar@{.>}@/^/[r]^c \ar@2{.>}@/^/[d]^{\ell}
        &\ar@2{.>}@/^/[r]^c&\\
        0,1&1,0&0,1&1,0&0,1&1,0&& }
  \caption{\textsf{Alice} leaves always  and \textsf{Bob} continues always.}\label{fig:AaBc}
}}
\end{figure} 
\item \textsf{Alice} continues always and \textsf{Bob} leaves always  (Figure~\ref{fig:AcBa}).
\begin{figure*}[hbt!]
    \centering
     \doublebox{\parbox{.85\textwidth}{
     \begin{center}
     \xymatrix{ *++[o][F]{\Al} \ar@{=>}@/^/[r]^c \ar@/^/[d]^{\ell} &*++[o][F]{\Be}
        \ar@/^/[r]^c \ar@{=>}@/^/[d]^{\ell} &*++[o][F]{\Al} \ar@{=>}@/^/[r]^c \ar@/^/[d]^{\ell}
        &*++[o][F]{\Be} \ar@/^/[r]^c \ar@{=>}@/^/[d]^{\ell} &*++[o][F]{\Al} \ar@{=>}@/^/[r]^c
        \ar@/^/[d]^{\ell} &*++[o][F]{\Be} \ar@{.>}@/^/[r]^c \ar@{=>}@/^/[d]^{\ell}
        &\ar@2{.>}@/^/[r]^c \ar@{.>}@/^/[d]^{\ell} &\ar@{.>}@/^/[r]^c&\\
        0,1&1,0&0,1&1,0&0,1&1,0&& }
    \end{center}
    \caption{\textsf{Alice} continues always and \textsf{Bob} continues always.}\label{fig:AcBa}
  }}
   \end{figure*}
\end{enumerate}

\begin{figure}[htb!]
  \centering
  \doublebox{\parbox{\textwidth}{
    \begin{displaymath}
      \xymatrix@C=10pt{
        &\ar@{.>}[r]& *++[o][F]{\Al} \ar@/^1pc/[rr]^c \ar@/^/[d]^{\ell} 
        &&*++[o][F]{\Be} \ar@/^1pc/[ll]^c \ar@/^/[d]^{\ell} \\
        &&0,1&&1,0
      }
      \qquad 
      \xymatrix@C=10pt{
        &\ar@{.>}[r]& *++[o][F]{\Al} \ar@{=>}@/^1pc/[rr]^c \ar@/^/[d]^{\ell} 
        &&*++[o][F]{\Be} \ar@/^1pc/[ll]^c \ar@{=>}@/^/[d]^{\ell} \\
        &&0,1&&1,0
      } 
      \qquad
      \xymatrix@C=10pt{
        &\ar@{.>}[r]& *++[o][F]{\Al} \ar@/^1pc/[rr]^c \ar@{=>}@/^/[d]^{\ell} 
        &&*++[o][F]{\Be} \ar@{=>}@/^1pc/[ll]^c \ar@/^/[d]^{\ell} \\
        &&0,1&&1,0
      }
\end{displaymath}
}}
\caption{The $0,1$ game and its equilibria seen compactly}
\label{fig:boucle}
\end{figure}
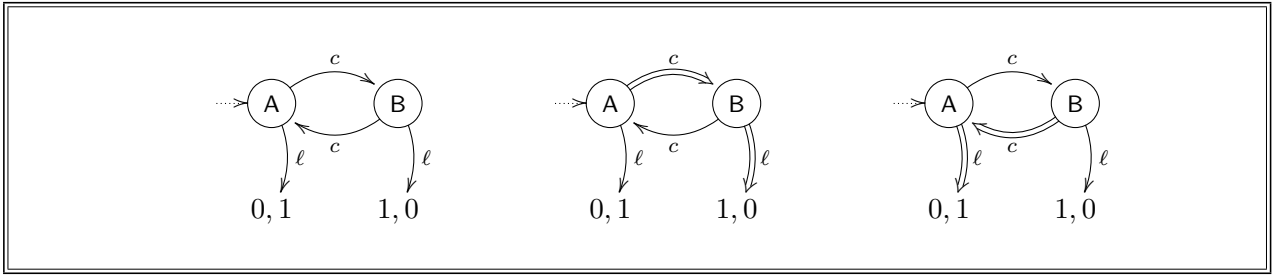

\begin{small}
  \paragraph{Justification} \emph{This paragraph can be skipped at first reading}.
  Let us see why and how this works.  Let us recall how we proceeded in the case of finite
  games.  Starting of end games, we examine a game position.  There are a player and two
  subgames of which we know at least an equilibrium: we choose as an equilibrium for the
  whole game the one that corresponds to the player making the choice of the best of the
  two equilibria.  Now let us go to infiniteness and let us show that the strategy
  profile where \textsf{Alice} always leaves and \textsf{Bob} always continues is an
  equilibrium.  Like in the case of the backward induction, we know the equilibria for the
  subgames.  In the case where it is \textsf{Bob}'s turn, what are the equilibria?  There
  are two, one is \textsf{Bob}'s abandon and in the other, where \textsf{Bob} continues, one
  has a sub-equilibrium in which \textsf{Alice} starts.  What is the equilibrium where
  \textsf{Alice} starts?  This is the same as this we are looking for, namely the strategy
  profile where \textsf{Alice} always leaves and \textsf{Bob} always continues.  Among the
  optimal strategy profiles available to \textsf{Bob}, which one is the best?  This where
  \textsf{Bob} continues, since \textsf{Alice} abandons and he wins $1\!\euro$, wherever if he
  would leave, he would  earn nothing.  Hence the good choice, which yields the equilibrium, is
  this where \textsf{Bob} continues, then \textsf{Alice} leaves always, then \textsf{Bob}
  continues always.  In the case of \textsf{Alice}'s turn, a similar reasoning shows that
  an equilibrium is this where \textsf{Alice} always leaves and \textsf{Bob} always continues.
  Let us summarize: We have shown that if one makes the hypothesis that the equilibrium is
  the strategy profile where \textsf{Alice} always leaves and \textsf{Bob} always continues
  then the equilibrium is the strategy profile where \textsf{Alice} always leaves and
  \textsf{Bob} always continues.  This a bit sophisticated explanation may be better
  understood on Figure~\ref{fig:boucle} where we have represented the game $0,1$ more
  compactly.  Are we cycling?  No!  Our reasoning is perfectly correct, because the
  hypothesis is on strict subgames.  Let us call it \emph{backward coinduction}: induction
  comes from the noun \emph{induction} and the prefix \emph{co}, ``associated'', and
  \emph{backward} insists on the analogy with backward induction.  Actually we have shown
  that the strategy profile where \textsf{Alice} leaves always and \textsf{Bob} continues
  always is an equilibrium all the game along.  We say that this is an ``invariant'' of the
  infinite game.
\end{small}

The same kind of reasoning applies to the dollar auction.  This is slightly more complex,
since the invariant is not the same strategy profile, but the same parametrized strategy
profile to take into account the fact that the involved numbers increase.  Then there are at
least two equilibria in the dollar auction: one for \textsf{Alice} continuing always and
\textsf{Bob} leaving always and the other for the other way around.

\section{Escalation is rational}
\label{sec:LescEstRati}
\hfill\parbox{12cm}{
Irrational exuberance has unduly escalated asset values.

\rightline{\emph{Alan Greenspan,}} \rightline{\textsf{Federal Reserve Board Chairman, on
    May 12 1996}} }

\bigskip

We know equilibria of the $0,1$ game, but how this may lead to escalation?  Assume that
our players or economic actors are rational, but have no memory and no ability to
reconsider their choice.  They forget immediately and learn nothing from past. They know
how to analyze a situation and understand their interest, but they do not know how to take
advantage from their experience, they do not see themselves reasoning and they have no reflection on
their own reasoning.  Moreover they do not take into account the marginal fees and they
do not see a moving society.  In
short they are introvert and forgetful.  At each step of the $0,1$ game, \textsf{Alice}
starts a new infinite $0,1$-game. She knows she has two possible rational strategies. The
first strategy consists in continuing today, tomorrow, the day after tomorrow and always,
assuming that \textsf{Bob} will leave always.  In this second strategy, she takes very
seriously the resolution of \textsf{Bob} to continue always. This is sometime called a
\emph{credible threat}, which is the attitude of Morris and Steel's friends\footnote{Among
  five pamphlet distributors, only Morris and Steel fought back MacDonald's.} in front of
MacDonald's.  \textsf{Alice} will continue if she thinks that she impressed enough
\textsf{Bob} to scare him and make him to abandon always.  This is MacDonald's attitude
all along the case against Morris and Steel.  ``This is not possible. They are going to
give up'' thought MacDonald's board.  If we are in a situation where no one takes
seriously the threat of the other, no threat is credible, we are in an escalation.

Let us recall two characteristics of escalation. Like in any sequential game, only
comparisons\footnote{In any case, using probabilities, assuming we know on what it
  applies, would not add anything of the prescription in successive choices.} are
pertinent.  Agents handle entities that are completely abstract: they do not know the
value of what they handle. They know only how entities are compared.  Moreover, agents are
faced at each step to two (or more) options equally rational, therefore the evolution of
the process is highly unpredictable.  At this level exogenous influences on the decision
process may happen.  Since \textsf{Alice} has no objective reason to choose between
``continuing'' or ``leaving'', she can take into consideration emotional
aspects\footnote{Daniel Kahneman, \emph{Thinking, Fast and Slow}, Farrar, Straus and
  Giroux, 2011.} or invoke other rational criteria.

\section{Escalation and cognitive psychology}
\label{sec:esc-psy}

\bigskip

It is worth wondering whether agents are really rational.  Take for this the view of
cognitive psychology as describe in Keith E. Stanovich\emph{What Intelligence Tests Miss:
  the psychology of rational thought}.\footnote{\textsf{Yale University Press 2010}.  One
  can also read Antonio Damasio \emph{Descartes' Error: Emotion, Reason, and the Human
    Brain}, Putnam, 1994} A rational agent owns a \emph{mindware} made of rules, knowledge
and procedures which he can retrieve in her memory and which she acquired by a mental
training and/or by education and which allows her to make decision and to solve problem.
Of course we assume that the mindware of the agents we consider contents coinductive
reasoning tools or all types of equivalent reasonings which allows conducting correct
deductions on infinite mathematical objects.\footnote{Stanovich seems to not be aware of
  coinduction and therefore considers escalation as dysrational in his mindware.
  Nevertheless his distinction between instrumental and epistemic rationality remains
  operational. }

There are two kinds of rationality from the coarser one to the finer one.  On
one side the instrumental rationality allows the agent to behave appropriately among the
world so that she gets what she wishes, using her physical and mental resources.  The
economists and the cognitive scientists refined the notion of wished goal to this of
\emph{expected utility}.  The \emph{epistemic rationality} lies above instrumental
rationality and interacts with it.  It allows the agent to confront her set of beliefs to
the effective structure of the world.  We would say that it makes the agent able to think
about her own way of reasoning.  More conventionally, we would say that the epistemic
rationality deals with what is true, whereas instrumental rationality deals with actions
to maximize aims.  The first form of rationality corresponds to algorithmic mind and the
second one to reflexive mind.  A reflexive mind is able to analyze how she reasons.

We claim that an escalating agent owns a correct algorithmic mind, but meanwhile she lacks
a reflexive mind which would allow her, by revising her belief, specifically her belief in an
infinite resource, to escape from the escalation spiral.  Indeed at the beginning such a
belief gives her dynamism for investing, but as the adage says ``Errare humanum est,
perseverare diabolicum''.  Therefore there is a time when a rational agent must understand
that a belief in an infinite resource leads her to a dead end and that her judgement must be
revised.  Doing so, early enough, the agent shows that she is really rational.

\section{The ubiquity of escalation }
  \hfill\parbox{12cm}{
\begin{quote}
She bought ostrich feathers, Chinese porcelain, and trunks; she borrowed from F\'elicit\'e,
from Madame Lefran\c cois, from the landlady at the Croix-Rouge, from everybody, no matter 
where.    With the money she at last received from Barneville she paid two bills; the other fifteen
hundred francs fell due. She renewed the bills, and thus it was continually.

\rightline{\emph{Gustave Flaubert}} \rightline{\textsf{Madame Bovary}}
\end{quote}}

\bigskip

Escalation is a very frequent phenomena as soon as participants are rational and consider
infinite resources.  In some cases this is what people look for, because this is a survival
condition, like in the case of evolutive biology. In his \emph{Red Queen} theory  Leigh
Van Valen describes the competition of two species and the survival condition of each
species that results, namely a continual adaptation to fight the challenge of the
species.  Escalation is not a drawback, but a positive quality that makes the species
perennial when it has a challenger.  In another hand, in economy, we know perfectly well
the consequence of escalation, namely speculative bubbles with huge devastations when
they blow up. 

\begin{figure}[htb]
  \centering
  \begin{center}
    \includegraphics[width=.9\textwidth]{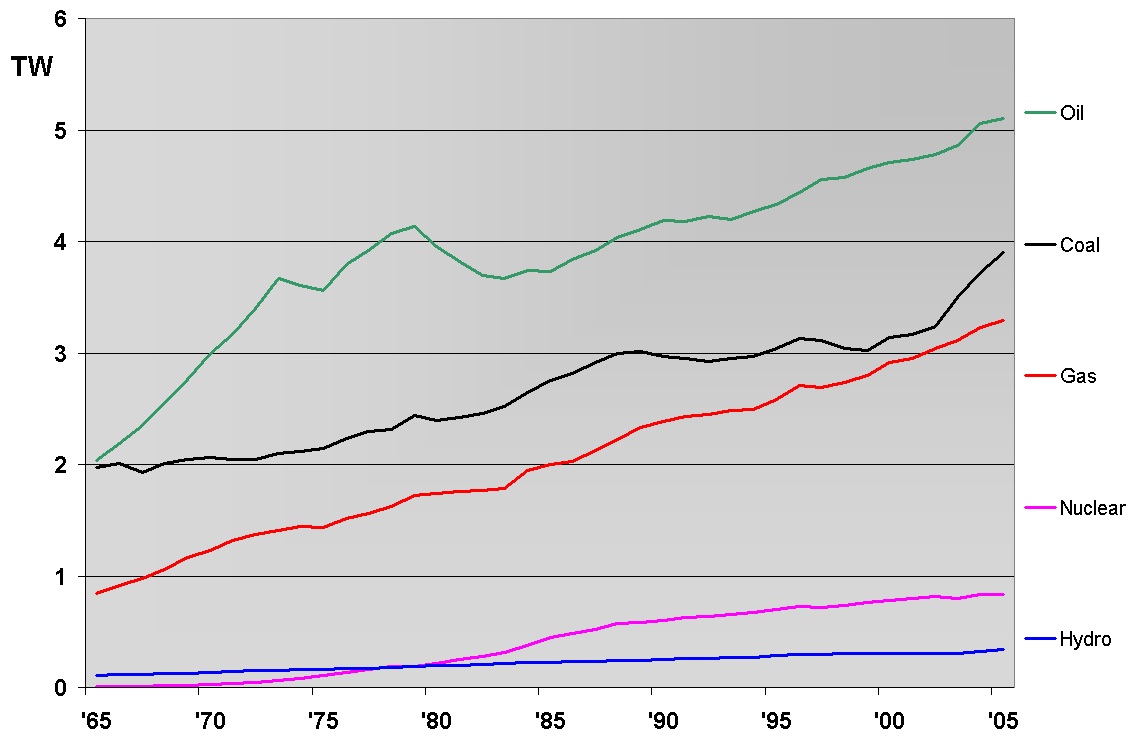}
  \end{center}
  
  \caption{Evolution of world energy  consumption  (source Wikipedia)}
\label{fig:cons}
\end{figure}

Development theory says that earth provides us  with a finite amount of fossil energies
and the sun light is a limited resource as well. Therefore the increasing of energy consumption (Figure~\ref{fig:cons})
is a frightening escalation.

In another field, escalation is a consubstantial component of war.  From the Bear Hall
Putsch to his suicide, through  \emph{Mein Kampf}  publication and Stalingrad battle,
Hitler's trajectory is an escalation.  Locally and in isolation, Hitler has displayed a
strategic rationality, i.e., an instrumental rationality.

\section{It is not possible to extrapolate}
\hfill\parbox{12cm}{
  \begin{quote}
One morning we set sail, with brains on fire,

And hearts swelled up with rancorous emotion,

Balancing, to the rhythm of its lyre,

Our infinite upon the finite ocean.
\end{quote}
}

\rightline{\vdots\hspace*{6.5cm}}

\hfill\parbox{11cm}{
  \begin{quote}
We want, this fire so burns our brain tissue,

To drown in the abyss -- heaven or hell,
who cares? 

Through the unknown, we'll find the new.
  \end{quote}
\rightline{\emph{Charles Baudelaire}}

\rightline{\textsf{Le voyage} \textsf{\textit{(Travel)}}}

}

\bigskip

\begin{figure}[hbp!] \label{fig:julia}
  \centering
  \includegraphics[width=.5\textwidth]{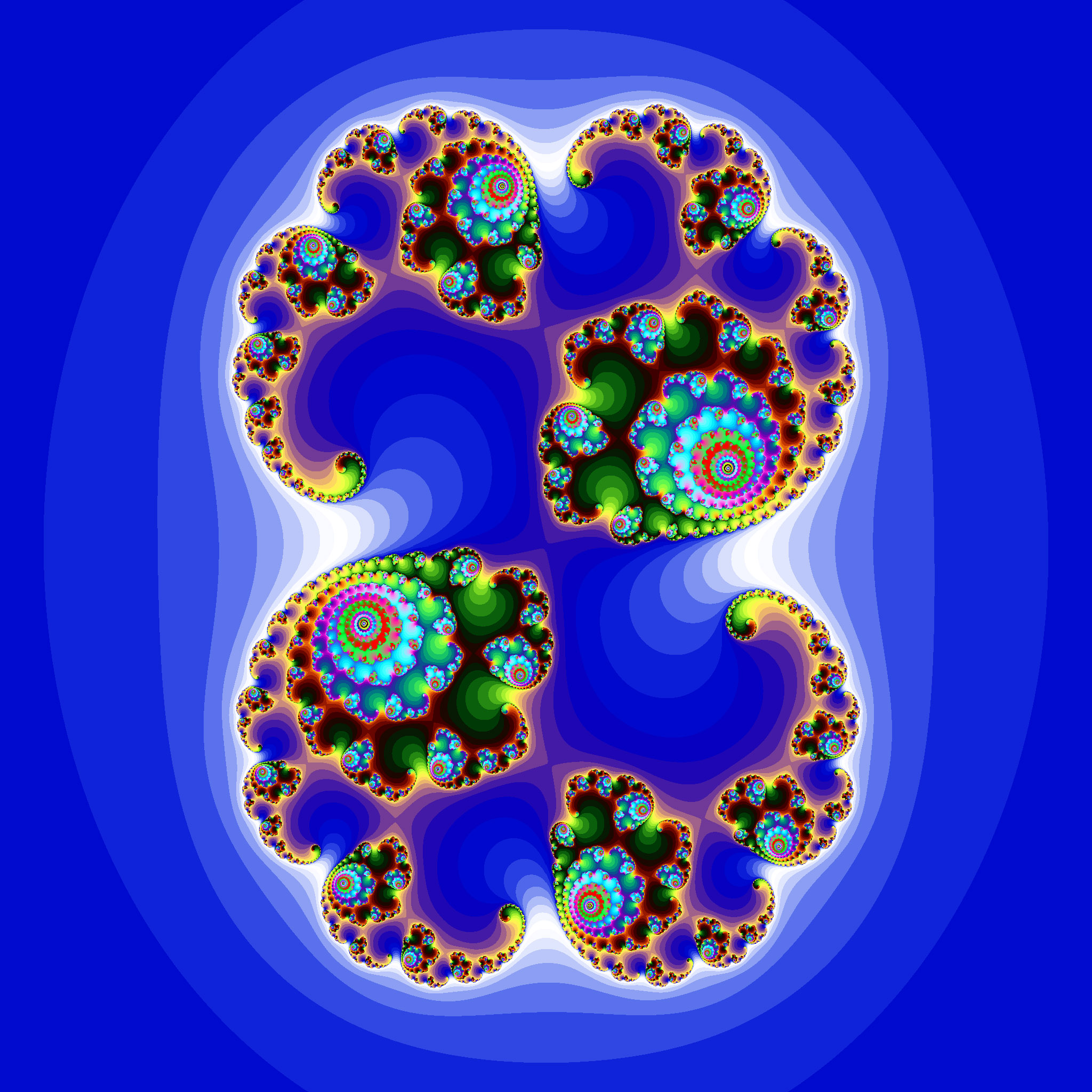}

\caption{The  Julia set, with a bound which is a tangentless curve}
\end{figure}

In Section~\ref{sec:back-on-0,1}, we saw that the reasoning mistake of Shubik and mostly
of his followers was to prune away an infinite branch of an infinite game to make finite
games, then to reason on finite games, then to extrapolate the results obtained on the
finite games to the infinite game.  We have seen that on the $0,1$ game this method cannot
work: according to the fact one prunes away at an even step or at an odd step the results
are different and there is no natural way to extrapolate.  In the case of the dollar
auction, the problem is more vicious since the results look consistent on the different
size of infinite games. However it may depend on the way the cut is done.  What is wrong
in the extrapolation has been well identified by the mathematician Wei\-er\-strass in
1871: what we know on the finite does not foretell what we can say on the infinite. More
precisely, he has proved that a well known and classical property of finite sums of
functions (to be differentiable everywhere and to be associated with smooth curves),
disappears for infinite sums of functions (which can be nowhere differentiable and be
associated with especially rough curves).  This result was a surprise when it has been
published, since great mathematicians before him admitted without proof the persistence at
infinity of the differentiability.  Moreover, this evidenced the existence of monster
curves without tangents, which have been systematically studied by Mandelbrot as fractals
(Figure~\ref{fig:julia}).
Actually this mistake on extrapolation goes back to Zeno of Elea, who stated that Achilles
would never overtake the tortoise. This led to negate motion.  Indeed Zeno extrapolated
the true result on one run that Achille does not overtake the tortoise, because he started
after the tortoise and reached the point where the tortoise started.  There are infinitely
many such runs.  Zeno is wrong when he extrapolates his result, true on one run, to the
limit of the infinite sequence of runs.  He should not conclude that Achille will not
overtake the tortoise.

Let us consider another example, namely a recent and easy to understand  mathematical
result,  founded on concepts known from Pythagoras and Euclid.   An \emph{arithmetic
progression} is a sequence obtained from some initial term by adding always the
same \emph{common difference}. The sequence $5, 8, 11, 14, 17, ... $ is
an arithmetic progression of common difference $3$.   There exists no infinite arithmetic
progression made only of primer numbers.\footnote{A prime number is a number divisible by
  the two numbers $1$ and itself.  If the origin of the progression is $n$ and its common
  difference is $d$, then the $(n+1)^{th}$ element is $n+ n\times d$ is clearly not prime,
  because it is divisible by $n$.}  Ben Green and Terence Tao have proved in 2004 that
there exist arbitrary long finite arithmetic progressions made only of prime
numbers.  This difficult result deserved Tao the Fields medal.  We face clearly a result
that is not extrapolable.   The very specific interest of this result is that the finite
case is incommensurably more difficult that the infinite case.  We were more used to the
opposite case, namely when the finite case is easier than the infinite case which
requires the subtle coinduction. 

\section{Conclusion}
\label{sec:conclusion}

By a precise analysis of the infinite, we have shown that agents involved in an escalation
are (instrumentally) rational.  Consequently, any approach that affirms too fast that they are
irrational\footnote{We do not deny they are party irrational. Indeed bewildered, they can
  invoke irrational arguments to raise perplexity. }  has missed the right argument based
on a coinductive analysis of infinite games.  The use of coinduction and more generally of
reasonings on infinite objects should be part of the new foundation of economics or as
cognitive psychology scientists would say, coinduction must be part of the mindware of the
economists.  In this framework, equilibrium will not be paired with stability.

On another hand the internal rationality of the agent should be made distinct from the
external rationality of the observer.  The agent who stipulates an infinite availability
of resource is introvert. She sees only her own short term interest and lacks of
``reflection''.  She is not able or she is not willing to imagine a global analysis,
whereas the observer sees immediately the behavioral aberration of the system.  Each one
has his own rationality and the points of view cannot be reconciled.


\pagebreak[2]

\end{document}